\documentclass[10pt,journal]{IEEEtran}
\usepackage{amsmath}
\usepackage{graphicx}
\usepackage{amsmath,amssymb,amsfonts}
\usepackage{algorithmic}
\usepackage{booktabs}
\usepackage{cite}
\usepackage{array}
\usepackage{url}
\usepackage{verbatim}
\usepackage{stfloats}
\usepackage{textcomp}

\usepackage{makecell}
\usepackage{threeparttable}
\usepackage{multirow}

\usepackage{float}
\usepackage{subfigure}
\usepackage{xspace}

\newcommand{\name}{MERIT\xspace
}
\def\BibTeX{{\rm B\kern-.05em{\sc i\kern-.025em b}\kern-.08em
    T\kern-.1667em\lower.7ex\hbox{E}\kern-.125emX}}
\begin{document}

\begin{sloppypar}

\title{MERIT: Multimodal Wearable Vital Sign Waveform Monitoring}

\author{Yongyang~Tang, Zhe~Chen,~\IEEEmembership{Member,~IEEE,} Ang~Li,~\IEEEmembership{Member,~IEEE,} Tianyue~Zheng,  Zheng~Lin, Jia~Xu, Pin Lv, Zhe Sun, and  Yue Gao,~\IEEEmembership{Fellow,~IEEE}}

\maketitle
\begin{abstract}
Cardiovascular disease (CVD) is the leading cause of death and premature mortality worldwide, with occupational environments significantly influencing CVD risk, underscoring the need for effective cardiac monitoring and early warning systems. 
Existing methods of monitoring vital signs require subjects to remain stationary, which is impractical for daily monitoring as individuals are often in motion.
To address this limitation, we propose MERIT, a multimodality-based wearable system designed for precise ECG waveform monitoring without movement restrictions.  
Daily activities, involving frequent arm movements, can significantly affect sensor data and complicate the reconstruction of accurate ECG signals. To mitigate motion impact and enhance ECG signal reconstruction, we introduce a deep independent component analysis (Deep-ICA) module and a multimodal fusion module. We conducted experiments with 15 subjects. Our results, compared with commercial wearable devices and existing methods, demonstrate that MERIT accurately reconstructs ECG waveforms during various office activities, offering a reliable solution for fine-grained cardiac monitoring in dynamic environments.
\end{abstract}


\section{Introduction}\label{sec:introduction}

Electrocardiography (ECG) is a crucial tool for diagnosing heart health, widely used in various health monitoring scenarios, including measuring heart rate, assessing heart rhythm, diagnosing cardiac abnormalities, emotion recognition, and biometric identification \cite{zhang2015occupation,li2024privacy}. Daily monitoring of physiological signals, such as ECG, is essential for both medical diagnosis and overall health management at home and in the office \cite{wu2022echohand}. Such continuous monitoring provides valuable information, enabling timely identification of health issues and facilitating appropriate therapeutic interventions. This proactive approach can significantly improve patient outcomes and quality of life.

\begin{figure}[ht]
\centering  
\subfigure[Stationary state]{
\label{fig:station}
\includegraphics[width=0.48\linewidth]{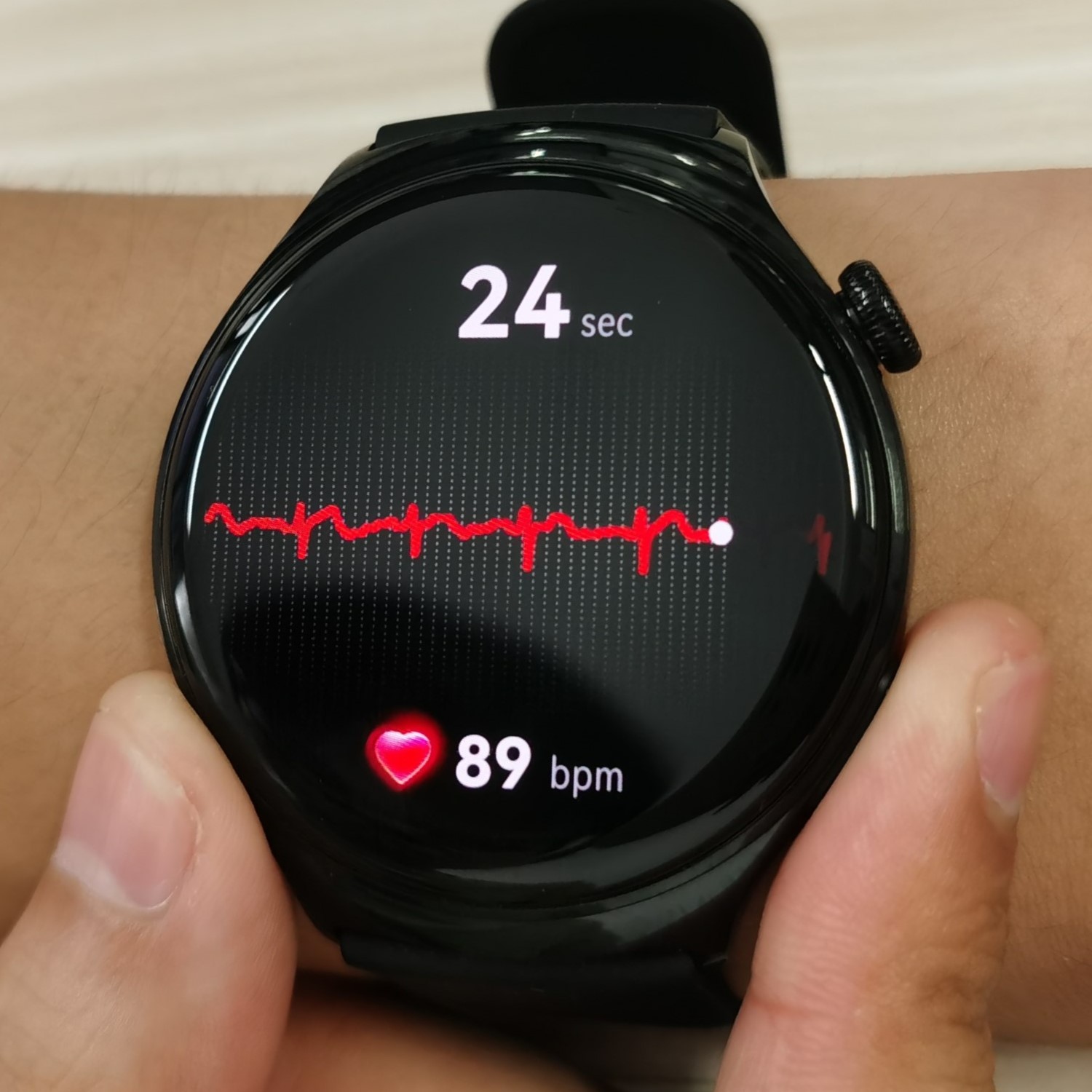}}
\subfigure[Motion state]{
\label{fig:motion}
\includegraphics[width=0.48\linewidth]{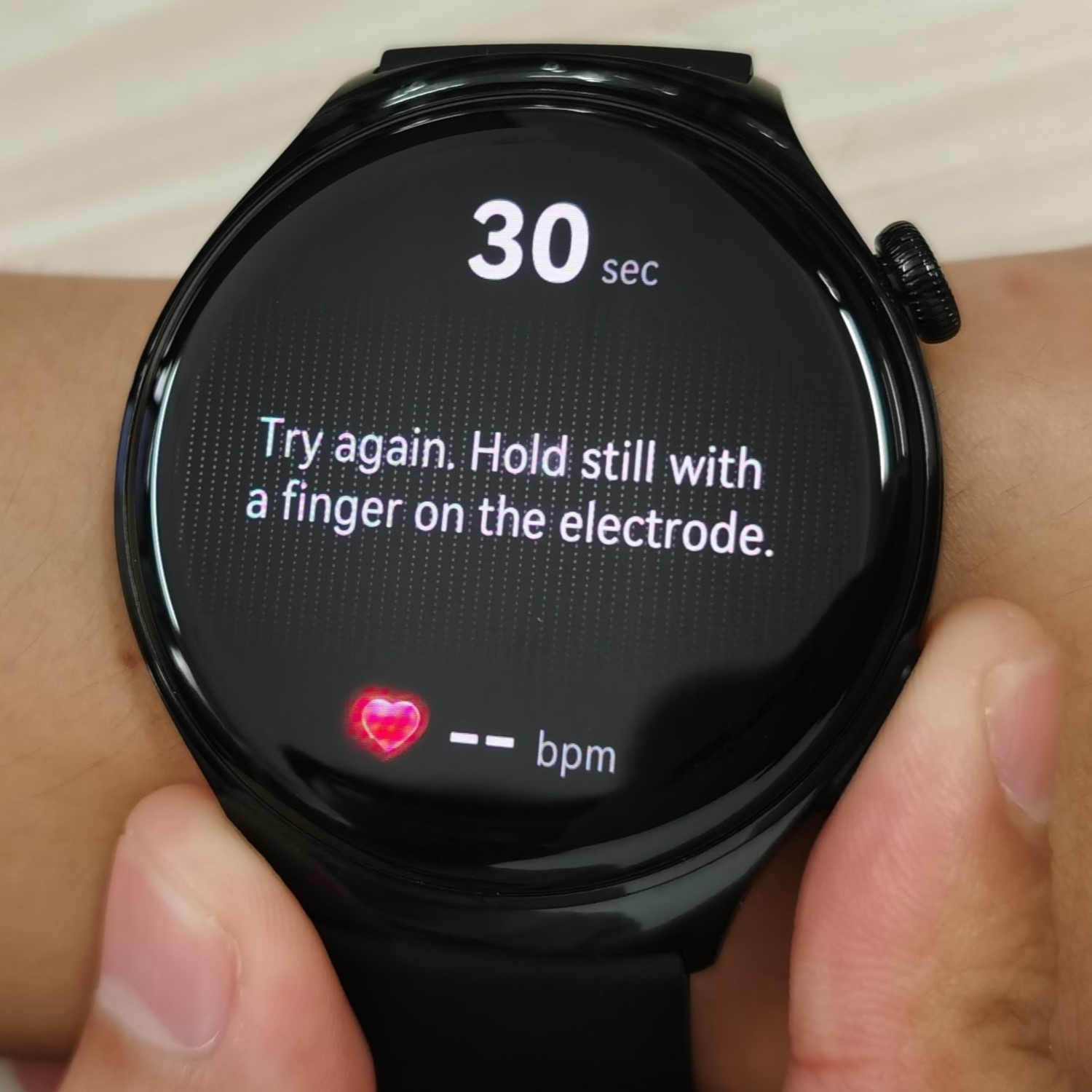}}
\caption{The effect of ECG acquisition by Huawei Watch 4 in different states}
\label{fig:watch}
\end{figure}

Currently, there are several commercial off-the-shelf (COTS) wearable devices that enable ECG monitoring, such as the Huawei Watch 4. This device integrates single-lead electrodes into the watch, allowing users to perform ECG monitoring in a specific position. However, as shown in Fig. \ref{fig:watch}, the Huawei Watch 4 is unable to perform ECG acquisition while the user is in motion. This limitation makes it unsuitable for continuous health monitoring tasks in dynamic office environments, where users are frequently active. A more robust solution is needed to address the challenges of accurate ECG monitoring during daily activities.

Numerous wearable devices and sensors have been developed for cardiac monitoring. For instance, VibCardiogram \cite{cao2022guard} utilizes a wearable inertial measurement unit (IMU) to recover ECG waveforms, offering a relatively flexible deployment setup. Additionally, various research initiatives have proposed non-contact methods for monitoring vital signs using wireless signals, which capture body micro-movements through reflected signals. Chen et al.~\cite{chen2024contactless} introduced a millimeter-wave radar system capable of observing cardiac activity and reconstructing the ECG without any physical contact. While these methods are more convenient than traditional electrode pads, they still require a fixed environment or the subject to remain relatively stationary. Previous studies on radar-based health monitoring during exercise \cite{chenMoViFiMotionrobustVital2021} can capture heart and chest vibrations to extract heartbeats, but these methods differ significantly from wearable ECG monitoring. Therefore, we designed a novel wearable vital signs monitoring solution \cite{li2024privacy}.

Noise in raw data can significantly impact the final results. Several studies have attempted to mitigate this issue using Independent Component Analysis (ICA) methods. ICA addresses blind signal separation problems, effectively isolating noise in waveforms from various input sources \cite{mcduff2023camera}. For example, \cite{poh2011advancements} employed ICA to separate data from different video channels, reducing the effects of light-wave artifacts and thereby obtaining blood volume pulse measurements. In office scenarios, frequent arm movements introduce additional noise into the data, including motion artifacts \cite{wu2024s}. Traditional linear ICA methods struggle to resolve noise problems caused by such motion \cite{liang2024vulseye}. Accurately identifying and eliminating these noise effects is critical for reliable monitoring results.

To better address these challenges, we introduce multiple sensors in the ECG signal recovery task. While the IMU performs well in stationary states, radar outperforms the IMU in motion states. To ensure optimal results in both scenarios, we propose a multimodal fusion module that integrates features from both sensor types. Additionally, due to noise introduced by motion \cite{liang2024ponziguard}, the traditional linear ICA method falls short. Therefore, we propose a Deep-ICA method, which leverages nonlinear ICA based on deep learning models~\cite{chen2021rf,lin2022channel} to reduce the impact of motion noise.

To summarize, the main contributions of this paper are as follows.
\begin{itemize}
\item[$\bullet$] We introduce multi-dimensional sensors for the task of recovering ECG waveforms and design a multimodal fusion module to enhance this process.
\item[$\bullet$] To alleviate the effects of wrist motion noise, we design the Deep-ICA module.
\item[$\bullet$] To validate the effectiveness of our proposed solution, we conducted extensive experiments. The results demonstrate that our method outperforms current wearable solutions and COTS devices.
\end{itemize}

\section{Background and Motivation}
\subsection{Vital Sign Sensing}

Clinicians try to use a variety of sensing methods to understand the state of a patient's heart \cite{park2020heartquake,zhang2015occupation,hyvarinen2000independent}. ECG is a commonly used technique for monitoring the heart. It records changes in the electrical signals generated during the contraction and diastole of the heart by attaching multiple electrodes to the skin of the body, providing information about the electrical characteristics of the heart. ECG is widely used in the diagnosis of heart disease, cardiovascular health monitoring, and clinical management, but they require adhesive gels and can cause skin irritation during recording \cite{kaplanberkaya2018survey}. These devices are also expensive and require specialized personnel to operate.

In recent studies, various sensory signals have been employed for vital sign measurements, including cameras, WiFi, IMU, and radar. For instance, SiNC \cite{speth2023noncontrastive} implements a non-contrastive unsupervised learning approach based on cameras for vital sign measurements, using non-RPPG-specific video data for training without ground truth vital signs. MultiSense \cite{zeng2020multisenseb} employs spectral analysis of the amplitude of Channel State Information retrieved from commercial WiFi devices to obtain respiration rate information, enabling simultaneous monitoring of multiple respiration rates. RF-ECG \cite{wang2023ecggrained} offers a fine-grained, non-contact cardiac monitoring solution by analyzing the characteristics of Ultra Wide Band (UWB) signals containing heartbeat and respiration data and developing a neural network mechanism to convert UWB signals into ECG signals. However, these solutions require the subject to be in a specific position or to remain relatively stationary, which does not adequately meet the needs of heart monitoring in dynamic office scenarios.

\subsection{Motivation}

In Section \ref{sec:secevaluation}, we observed that the IMU method can effectively perform the ECG recovery task when the subject is stationary. However, during motion, the irregular arm movements pose a challenge. The sensor data from the IMU, collected along three axes (x, y, z), captures vibrations that are irregularly distributed across these axes. Given the IMU’s high sensitivity to motion, the desired vibration signals are often masked by the motion artifacts, making it difficult to extract the necessary information.
To overcome this limitation, we integrate radar sensors into the wearable devices. Unlike the IMU, radar sensors measure the relative distance between the radar and the wrist’s blood vessels, maintaining this distance within a controllable range even during motion. This characteristic allows the radar to outperform the IMU in capturing vibration signals during movement. Conversely, due to the radar’s lower sensitivity to vibrations compared to the IMU, the IMU performs better when the subject is stationary. To leverage the strengths of both sensors across different states, we propose a multimodal fusion module that integrates the features from both IMU and radar data, thereby enhancing overall performance.
Additionally, arm movements introduce non-linear noise that overwhelms the vibration signals needed for accurate ECG monitoring. Traditional linear ICA methods are insufficient to separate this noise from the original data effectively. Therefore, we propose the Deep-ICA method, which employs a non-linear ICA approach to mitigate the impact of motion noise, significantly improving performance in dynamic conditions.

\section{System Design}
\subsection{Overview}

\begin{figure*}[t!]
    \centering
    \vspace{1pt}
    \includegraphics[width=\linewidth]{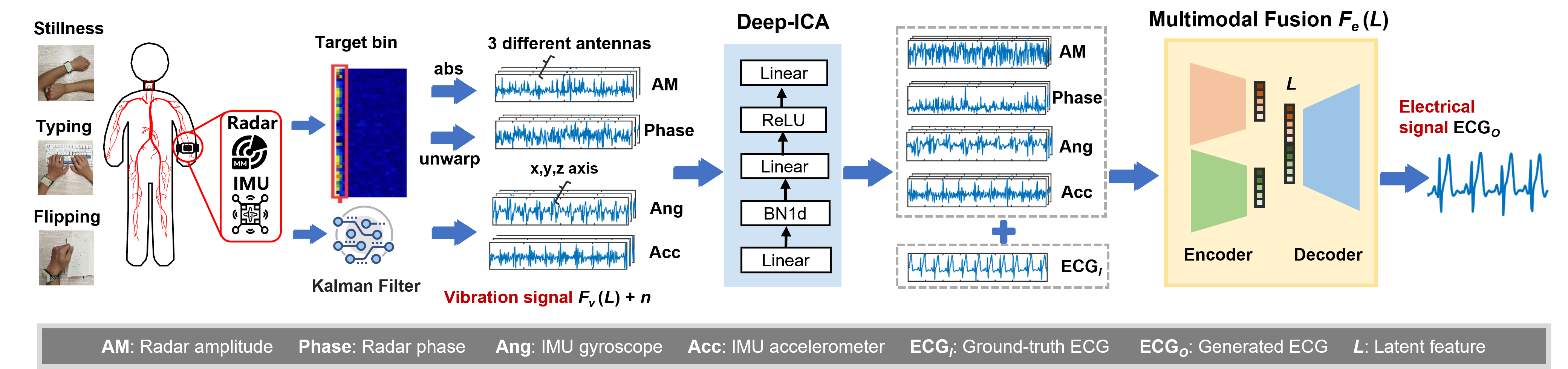}
    \caption{Overall design of \name.}
    \label{fig:demo}
     
\end{figure*}

In this section, we present the design of \name, which is composed of radar and IMU sensors, a deep independent component analysis (Deep-ICA), and a multimodal fusion module, as illustrated in Figure~\ref{fig:demo}. Before detailing our design, we model the process by which cardiac vibrations impact radar and IMU sensors (in Section~\ref{sec:define}). The Deep-ICA component, based on neural networks, is proposed to separate human motion from cardiac vibrations (in Section~\ref{sec:ICA}). Additionally, the multimodal fusion module is designed to integrate features from both IMU and radar sensors to accurately reconstruct ECG signals (in Section~\ref{sec:fusion}).

\subsection{Parameter Definition}\label{sec:define}

The most common method for acquiring electrical signals is through the use of an ECG device, which typically requires the attachment of electrodes to the body. However, non-contact measurement of electrical signals presents significant challenges. Recent studies \cite{park2020heartquake, chen2024contactless, cao2022guard} have demonstrated a strong correlation between electrical and mechanical cardiac activity. While both cardiogenic body vibrations and ECG signals represent the same cardiac events, they do so in different dimensions. The primary mechanism underlying cardiogenic body vibrations is explained by Newton’s Third Law of Motion, which states that for every action, there is an equal and opposite reaction. During cardiac systole and diastole, the body experiences periodic retractions due to the movement of large volumes of blood, causing vibrations in the wrist vessels. This phenomenon provides a theoretical foundation for deriving ECG signals from indirect measurements of these vibrations.
We hypothesize the existence of a latent feature, denoted as $L$, which serves as an intermediary between the electrical and vibration signals. This latent feature can be transformed into the electrical signal $F_e(L)$ and the vibration signal $F_v(L)$. Additionally, the vibration data collected by the sensor includes noise, denoted as $n$. Thus, the vibration signal collected by our sensor can be expressed as $V = F_v(L) + n$, while the electrical signal is expressed as $E = F_e(L)$. Given our objective to convert the vibration signal into an electrical signal, our task is to achieve this transformation:
\begin{equation}\begin{aligned} E = (F_v^{-1}(V - n)). \end{aligned}\end{equation}

\subsection{Deep ICA Module}\label{sec:ICA}
Blood vessel vibration signals exhibit numerous unpredictable patterns, and the mixing of these signals is evident in the detected readings from various axes in the IMU data, as well as from the target bin in the radar data. The identification of vibration signals from mixed sources falls within the broad scope of blind source separation. If the sources are independent, non-Gaussian and linearly combined, we can effectively achieve blind source separation through a technique known as ICA. 
Inspired by the work of \cite{khemakhemICEBeeMIdentifiableConditional2020}, we apply ICA to process the raw data. For the IMU data, we use different axes as inputs. In the case of radar data, previous ICA measurements are typically conducted on the human chest, usually with different bins as inputs. We select the target bin with different antennas as the input data for our model. This choice is made because, if we use both the target bin and the subsequent bin, the small diameter of the blood vessels in the wrist results in the subsequent bin lacking information about the blood vessel vibrations, thereby introducing additional noise into the data.

ICA typically uses all sources as inputs. We address the nonlinear ICA problem by leveraging a neural network model that combines an Energy-Based Model (EBM) with nonlinear ICA theory. We collect a dataset of observations of tuples $(\mathbf{x}, \mathbf{y})$, where $\mathbf{x} \in \mathcal{\mathcal { X }} \subset \mathbb{R}^{d_{x}}$ is the main variable of interest, also called the dependent variable, and $\mathbf{y} \in \mathcal{Y} \subset \mathbb{R}^{d_{y}}$ is an auxiliary variable also called the conditioning variable. We use the index of the data as the $\mathbf{y}$ when processing the signal data. Consider two feature extractors $\mathbf{f}_{\boldsymbol{\theta}}(\mathbf{x}) \in \mathbb{R}^{d_{z}}$ and $\mathbf{g}_{\boldsymbol{\theta}}(\mathbf{y}) \in \mathbb{R}^{d_{z}}$ , which we parameterize by neural networks, and $\boldsymbol{\theta}$ is the vector of weights and biases. To alleviate notations, we will drop $\boldsymbol{\theta}$ when it's clear which quantities we refer to. These feature extractors are used to define the conditional energy function:
\begin{equation}
\mathcal{E}_{\boldsymbol{\theta}}(\mathbf{x} \mid \mathbf{y})=\mathbf{f}_{\boldsymbol{\theta}}(\mathbf{x})^{T} \mathbf{g}_{\boldsymbol{\theta}}(\mathbf{y}).
\end{equation}

The parameter $\boldsymbol{\theta}$ lives in the space $\Theta$ which is defined such that the normalizing constant Equation \ref{zconstant} is finite. Our family of conditional energy-based models has the form:
\begin{equation}
p_{\boldsymbol{\theta}}(\mathbf{x} \mid \mathbf{y})=\frac{\exp \left(-\mathbf{f}_{\boldsymbol{\theta}}(\mathbf{x})^{T} \mathbf{g}_{\boldsymbol{\theta}}(\mathbf{y})\right)}{Z(\mathbf{y} ; \boldsymbol{\theta})},
\end{equation}
\begin{equation}\label{zconstant}
Z(\mathbf{y} ; \boldsymbol{\theta})=   \int_{\mathcal{X}} \exp \left(-\mathcal{E}_{\boldsymbol{\theta}}(\mathbf{x} \mid \mathbf{y})\right) \mathrm{d} \mathbf{x}<\infty.
\end{equation}

The extractor $\mathbf{f}_{\boldsymbol{\theta}}$ consists of a linear layer, a normalization layer, and an activation function. The model comprises three linear layers. A normalization layer has been added to both the first and second linear layers. An activation function has been incorporated into the second and third linear layers. The normalization layer is BatchNorm1d and the activation function is LeakyReLU. The normalization layer is BatchNorm1d and the activation function is LeakyReLU. The extractor ${\mathbf{g}_{\boldsymbol{\theta}}}$ denotes the $d_{z}\times d_{y}$ learnable matrix.

Suppose that the wireless signal observed $\mathbf{x}\in\mathbb{R}^{d_x}$, arises from the nonlinear transformation $h$, which links the heart rate signal to the noise signal $\mathbf{z}\in\mathbb{R}^{d_z}$. We assume that the distribution of $z$ is conditional to the auxiliary variable $\mathbf{y}\in\mathbb{R}^{d_y}$, which is presented $\mathbf{z}\sim p(\mathbf{z}|\mathbf{y})$, $\mathbf{x}=\mathbf{h}(\mathbf{z})$. Here we assume $d_x=d_z=d$. From the proof in \cite{khemakhemICEBeeMIdentifiableConditional2020} we can assume that the density of the observed wireless signal has the following form:
\begin{equation}
p(\mathbf{z}|\mathbf{y})=\mu(\mathbf{z})e^{\sum_{i=1}^d\mathbf{z}_i\mathbf{T}_i(z_i)^T\boldsymbol{\lambda}_i(\mathbf{y}){-\Gamma(\mathbf{y})}}.
\end{equation}

Here, $\mu(\mathbf{z})$ is the base and $\Gamma(\mathbf{y})$ is the conditional normalization constant.  The exponential term is a cross-component factor in this exponential family, and the sufficient statistic $\mathbf{T}$ consists of $d$ functions, each of which is a function of a component $z_i$ of the Variation signal and the noise signal $\mathbf{z}$. The above equations together define a non-parametric model with parameters $(\mathbf{h},\mathbf{T},\boldsymbol{\lambda},\mu)$. For the special case $\mu(\mathbf{z})=\prod_i\mu_i(z_i)$, The distribution of $\mathbf{z}$ is factorized in each dimension, and the components $z_i$ are independent. The generative model then obtains the nonlinear ICA model \cite{khemakhem2020variationala}.

During the model training phase, the conditional Flow Contrastive Estimation \cite{gao2020flow} technique was implemented to enhance the model's parameters. The form of the flow model is presented $x=g_\alpha(z)$,$~z\sim q_0(z)$.
Where $q_0$ is a known distribution of random noise. $g_{\alpha}$ is a combination of invertible sequences of transformations where the logarithmic determinant of the Jacobian determinant of the transformation can be explicitly obtained. Let $q_{\alpha}(x)$ be the probability density of the model given the data point $x$ and the parameter $\alpha$. Then $q_{\alpha}(x)$ can be expressed as the probability density in the case of a change in variables
\begin{equation}
q_{\alpha}(x)=q_{0}(g_{\alpha}^{-1}(x))|\det(\partial g_{\alpha}^{-1}(x)/\partial x)|.
\end{equation}

The training process involves iterative updates of both the EBM and the flow model. In this process, the flow model adapts to a more robust contrast distribution or a more capable EBM-trained adversary through a GAN-like parameter estimation scheme, where $p_{\theta}$ and $q_{\alpha}$ engage in a minimax game with a uniform value function. Specifically, FCE learns the parameters of the EBM density $p_{\theta}$ by performing a surrogate classification task. This task involves generating noise from the distribution $q_{\alpha}$, which is also parameterized by the flow model, followed by logistic regression to distinguish between genuine data samples and noise samples. The objective function is expressed as a simple logarithmic probability:
\begin{equation}
\begin{aligned}
    \mathcal{J}_{\mathrm{CFCE}}(\boldsymbol{\theta},\boldsymbol{\alpha})=\mathbb{E}_{p_{\mathrm{dala}(\mathbf{x},\mathbf{y})}}\log\frac{p_{\boldsymbol{\theta}}(\mathbf{x}|\mathbf{y})}{q_{\boldsymbol{\alpha}}(\mathbf{x},\mathbf{y})+p_{\boldsymbol{\theta}}(\mathbf{x}|\mathbf{y})}+\\\mathbb{E}_{q_{\boldsymbol{\alpha}}(\mathbf{x},\mathbf{y})}\log\frac{q_{\boldsymbol{\alpha}}(\mathbf{x},\mathbf{y})}{q_{\boldsymbol{\alpha}}(\mathbf{x},\mathbf{y})+p_{\boldsymbol{\theta}}(\mathbf{x}|\mathbf{y})}.
\end{aligned}
\end{equation}

\begin{figure}
\centering  
\subfigure[Deep-ICA input and ECG.]{
\label{fig:rawphase}
\includegraphics[width=0.48\linewidth,trim={ 0 0 1.3cm 0},clip]{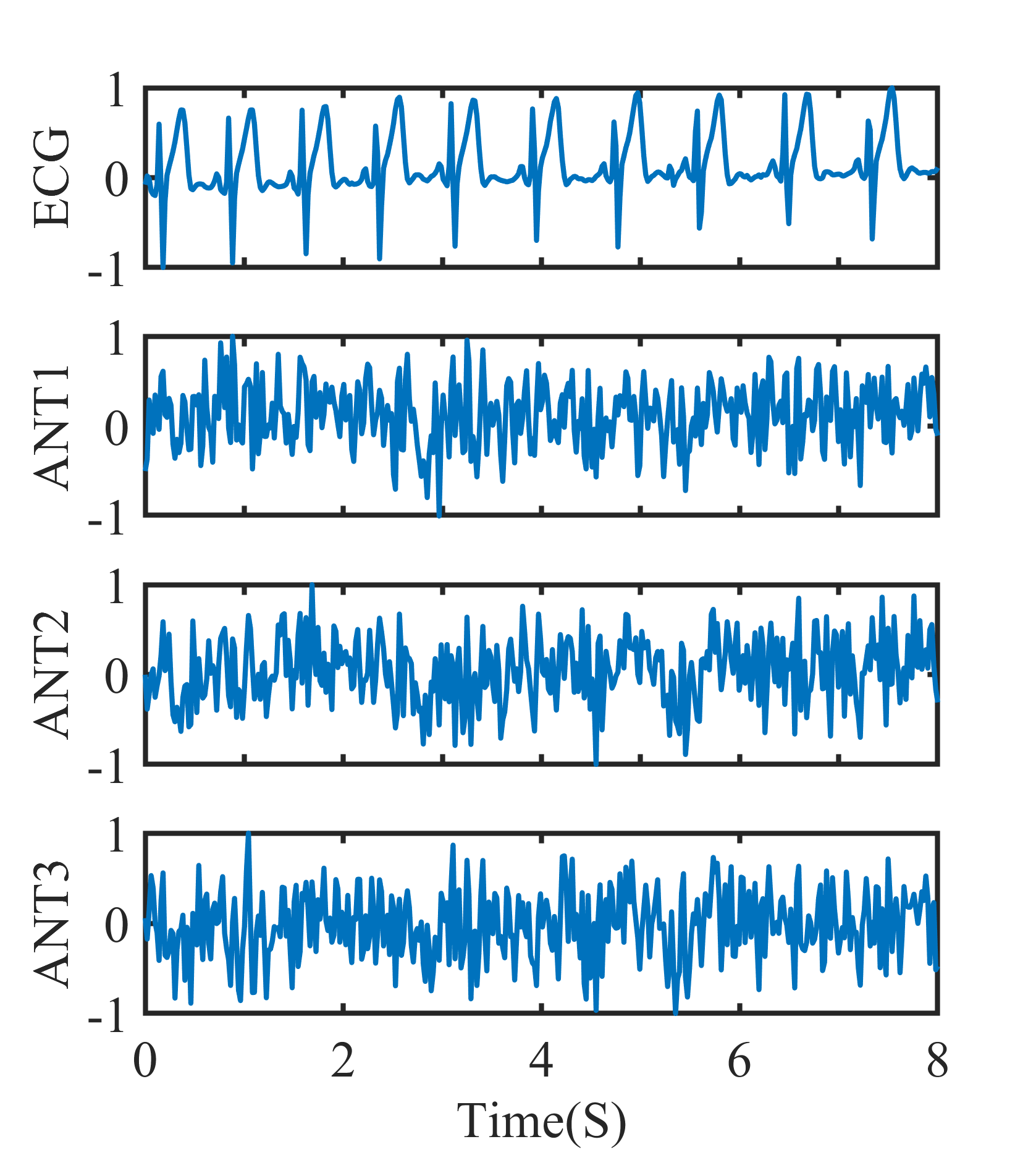}}
\subfigure[Deep-ICA output and ECG.]{
\label{fig:ICAphase}
\includegraphics[width=0.48\linewidth,trim={ 0 0 1.3cm 0},clip]{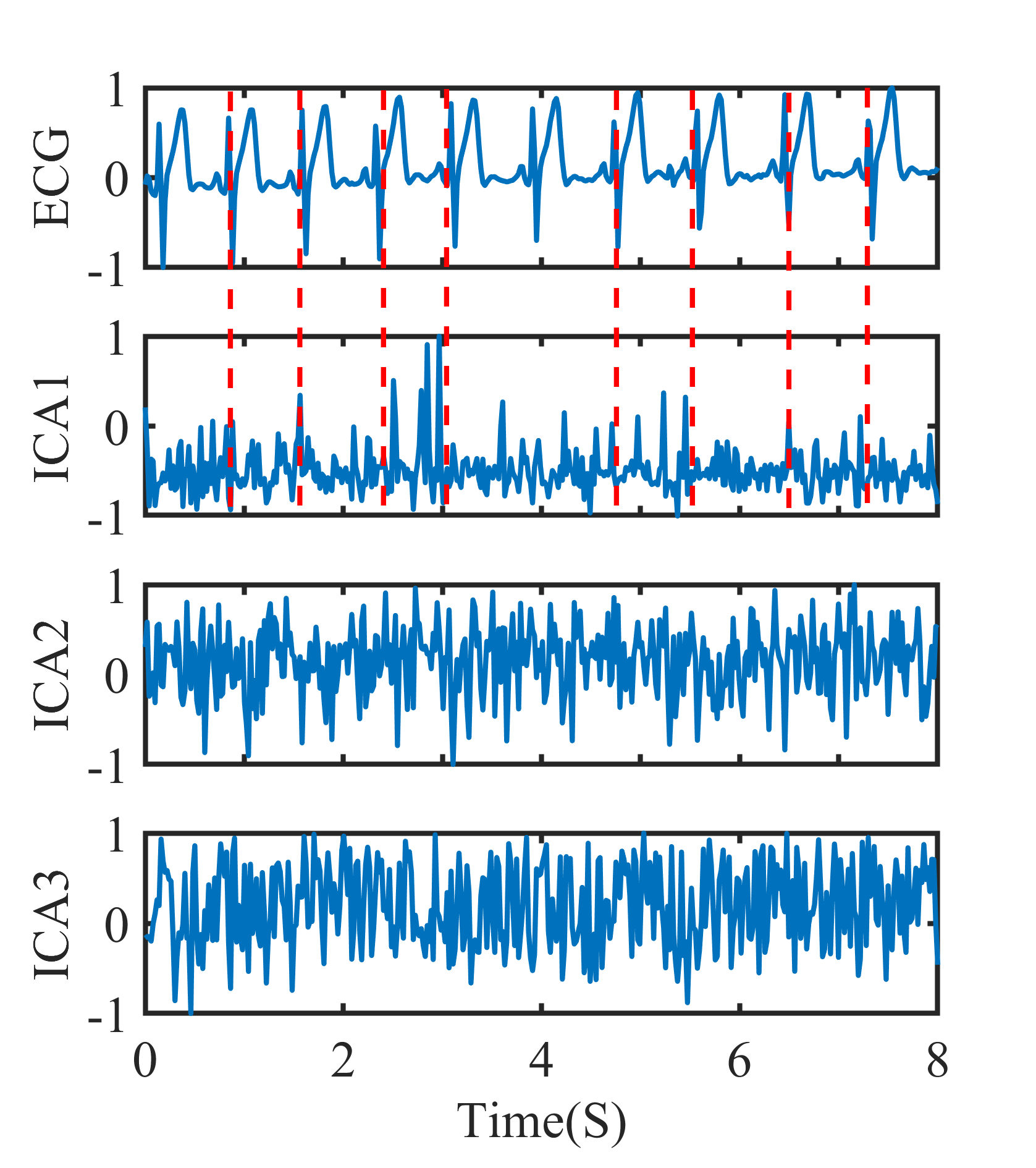}}
\caption{Input of Deep-ICA, correlation of Deep-ICA output with ECG signal}
\label{fig:ICAcomp}
\end{figure}

The target function is minimized with respect to $\theta$ and maximized with respect to $\alpha$. This approach has the added benefit of enhancing flow consistency. Here, expectation $\mathbb{E}_{p_\text{data}}$ represents an approximation of the observation sample expectation, while expectation $\mathbb{E}_{q_{\boldsymbol{\alpha}}}$ approximates the noise sample expectation. Additionally, we use conditional densities $p_{\boldsymbol{\theta}}(\mathbf{x}|\mathbf{y})$, which enable the incorporation of temporally sequential features into the model, leading to a more accurate estimation of the EBM. In this context, the conditioning variable $\mathbf{y}$ is indexed in discrete time, allowing the index to be sampled from a uniform distribution and the observations to be sampled using the conditional flow. The flow density can be expressed as $q_{\boldsymbol{\alpha}}(\mathbf{x},\mathbf{y})=p(\mathbf{y})q_{\boldsymbol{\alpha}}(\mathbf{x}|\mathbf{y})$.
In Fig. \ref{fig:ICAcomp}, we can see that the input data is effectively separated from the noise after passing through the Deep-ICA module. The separated model also corresponds to the ECG waveform to a higher degree.

\begin{figure*}
    \centering
    \includegraphics[width=0.9\linewidth]{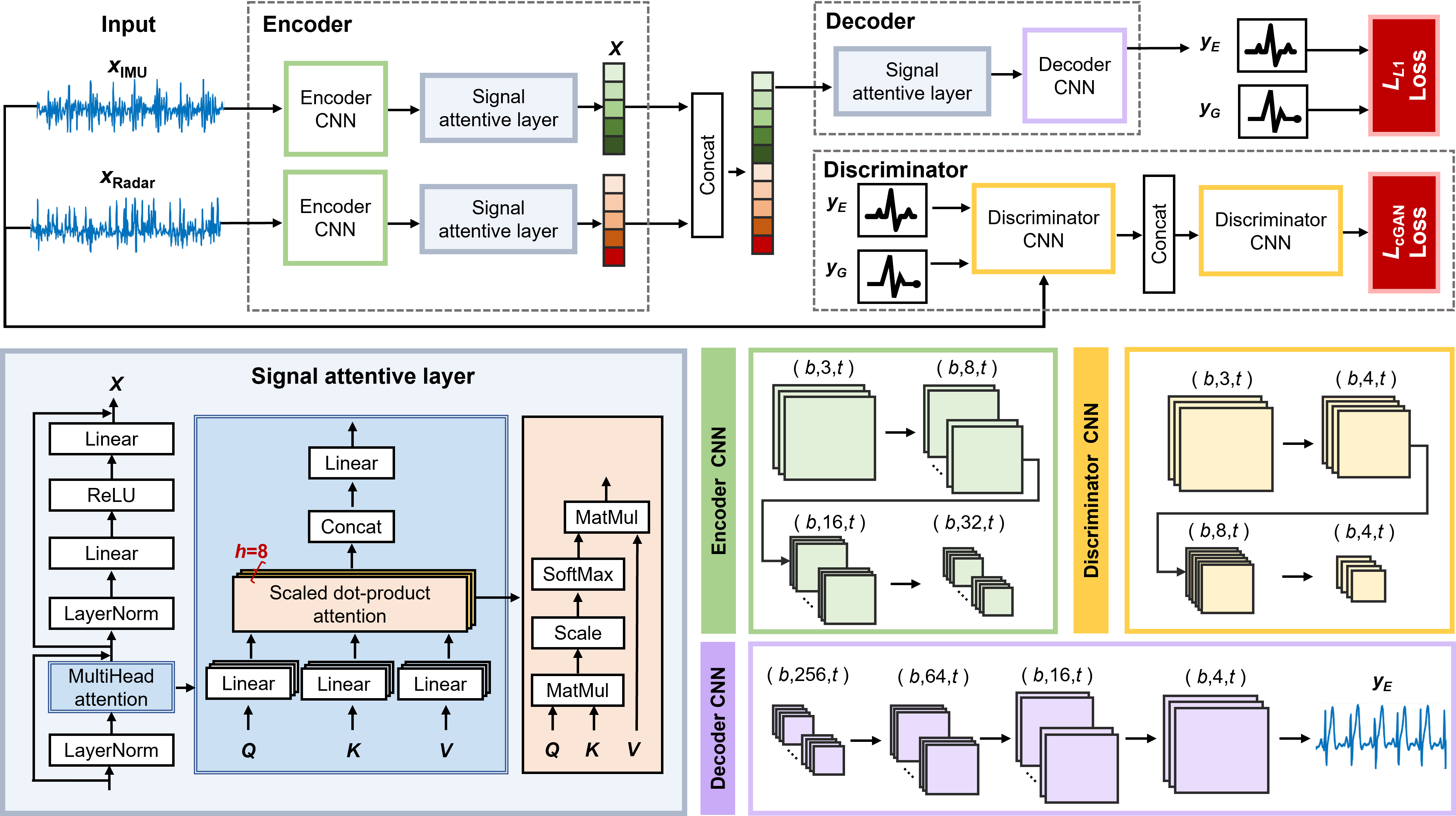}
    \caption{Multimodal fusion module structure}
    \label{fig:Reconstruction} 
\end{figure*}
\subsection{Multimodal Fusion Module}\label{sec:fusion}
The multimodal fusion module is designed to recover fine-grained ECG signals by mitigating the effects of small motion artifacts. This module utilizes a conditional GAN structure to extract features from various signals and reconstruct them as ECG signals. The generator $G$ is trained to produce outputs that are indistinguishable from real signals, while the discriminator $D$ is trained to differentiate between the generator's ``fakes'' and authentic signals. Fig. \ref{fig:Reconstruction} illustrates the module's architecture, which includes an adversarial network with both a generator and a discriminator. The generator features an encoder-decoder configuration, enabling effective recovery of ECG waveforms from time-series data such as IMU and radar. In this context, $b$ represents the batch size, $t$ denotes the length of the input data series, $y_E$ is the recovered ECG signal, and $y_G$ is the ground truth signal.

\textbf{Generator}
We employ an encoder-decoder architecture to reconstruct the vital signs waveform from IMU and radar signals. The model combines features extracted from the IMU data with the radar data by the encoder. As our data is basically 1D time series data, we use Conv1d as the CNN layer. The encoder structure includes both a CNN layer and a signal attentive layer. The output of the CNN goes through a normalization layer.

Fig. \ref{fig:Reconstruction} illustrates the structure of the signal-attentive layer. The preceding layer of the network first undergoes LayerNorm, followed sequentially by the self-attention layer, a linear layer, and an activation function. This layered structure may lead to gradients approaching zero, known as the vanishing gradient problem. To address this, we incorporate residual connections in both the self-attention and linear layers, which add the residual input values directly to the output of these components. To merge the representations from the two modalities, we concatenate the latent vector output from the encoder \cite{kim2021vilta}. The decoder includes a CNN layer. Additionally, we apply dropout to introduce noise into the network, enhancing its ability to learn diverse distributions. Finally, the generator produces the estimated ECG signal $y_E$.

\textbf{Discriminator}
Given the correspondence between our inputs and outputs, we employ a conditional GAN to guide the model in learning this relationship. It is feasible to preprocess the input radar and IMU data to generate the corresponding ECG signal. We then use the original radar and IMU data, along with the generated ECG signal data, as inputs to the discriminator. The loss function for the conditional GAN is:
\begin{equation}
\begin{aligned}
    \mathcal{L}_{cGAN}(G,D)=\mathbb{E}_{I_\text{Radar},I_\text{IMU},I_\text{ECG}}[\log D(I_\text{Radar},I_\text{IMU},I_\text{ECG})]+\\\\\mathbb{E}_{I_\text{Radar},I_\text{IMU},I_\text{ECG}}[\log(1-D(I_\text{Radar},I_\text{IMU},G(I_\text{Radar},I_\text{IMU})))].
\end{aligned}
\end{equation}

Here, $I_\text{Radar}$ is the radar data fed to the model. $I_\text{IMU}$ is the IMU data. $I_\text{ECG}$ represents the ground truth of the ECG signal. In previous methods \cite{pathak2016context}, the L2 distance would be used as a loss function to optimize between the generated data and the target data. However, research conducted by \cite{isola2017imagetoimage,wang2023ecggrained} has demonstrated that utilizing L1 distance as the loss function assists in mitigating signal blurring in comparison to L2 distance. So, we utilized the L1 distance as a loss function:
 
\begin{equation}
    \mathcal{L}_{L_1}(G)=\mathbb{E}_{I_\text{Radar},I_\text{IMU},O_\text{ECG}}\left[\|I_\text{ECG}-G(I_\text{Radar},I_\text{IMU})\|_1\right].
\end{equation}

Our final objective is:
\begin{equation}\label{eq:ganloss}
    \mathcal{L}=\mathcal{L}_{cGAN}+\alpha\mathcal{L}_{L_1}.
\end{equation}

Here, parameter $\alpha$ is an arbitrary constant used to balance the ratio between the two losses. Section 4, we find experimentally that a value of 9 for $\alpha$ d works best. After training with the discriminator D, the generator G can precisely translate radar and IMU data into ECG signal waveforms.

\section{Implementation} \label{sec:evaluation}
In this section, we describe the hardware part of the experimental equipment, the software to be used and the way the data was collected.
\subsection{Hardware}
Since there is no off-the-shelf multimodal wearable device, we designed a multimodal wearable device using the Infineon BGT60TR13C radar module and the ICM-42607-P IMU module. The BGT60TR13C is a 60 GHz radar sensor with an integrated antenna that is small enough to be integrated into our devices. It has a bandwidth of 5.5 GHz and supports FMCW modulation. It has one transmit antenna and three receive antennas. The ICM-42670-P is a high-performance 6-axis microelectromechanical system MotionTracking device that combines a 3-axis gyroscope and a 3-axis accelerometer. For ECG devices we use the vernier EKG Sensor, which records the electrical signals generated by heart or muscle contractions. It can output a standard 3-lead ECG tracing. We place the device on the user's wrist for collection. We used data collected by Vernier Go Direct's EKG Sensor devices as ground truth.

\begin{figure}[h]
	\centering
	\begin{minipage}{0.48\linewidth}
		\centering
		\includegraphics[width=\linewidth]{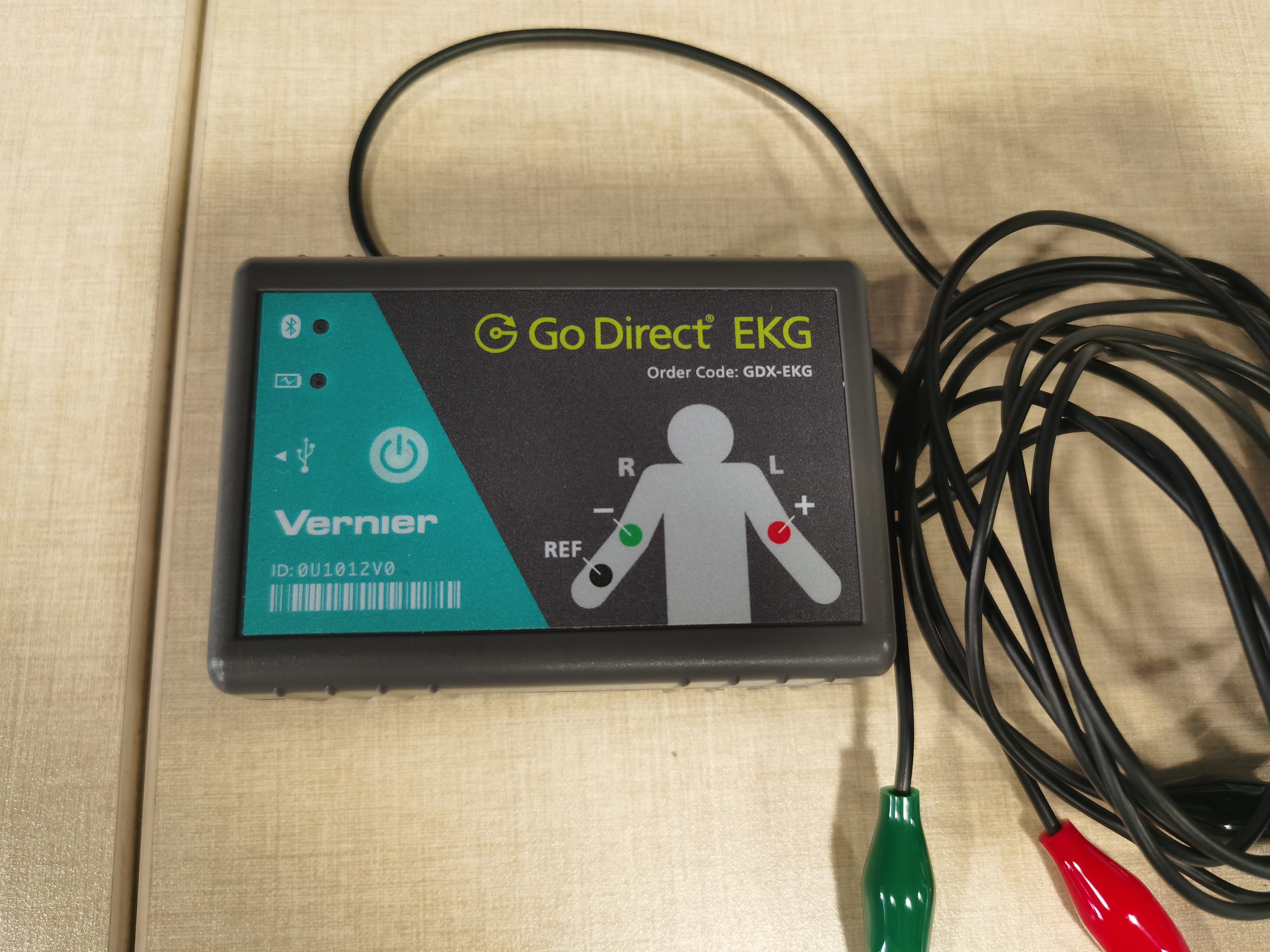}
		\caption{ECG device}
		\label{fig:ECGdevice}
         
	\end{minipage}
	\begin{minipage}{0.48\linewidth}
		\centering
		\includegraphics[width=\linewidth]{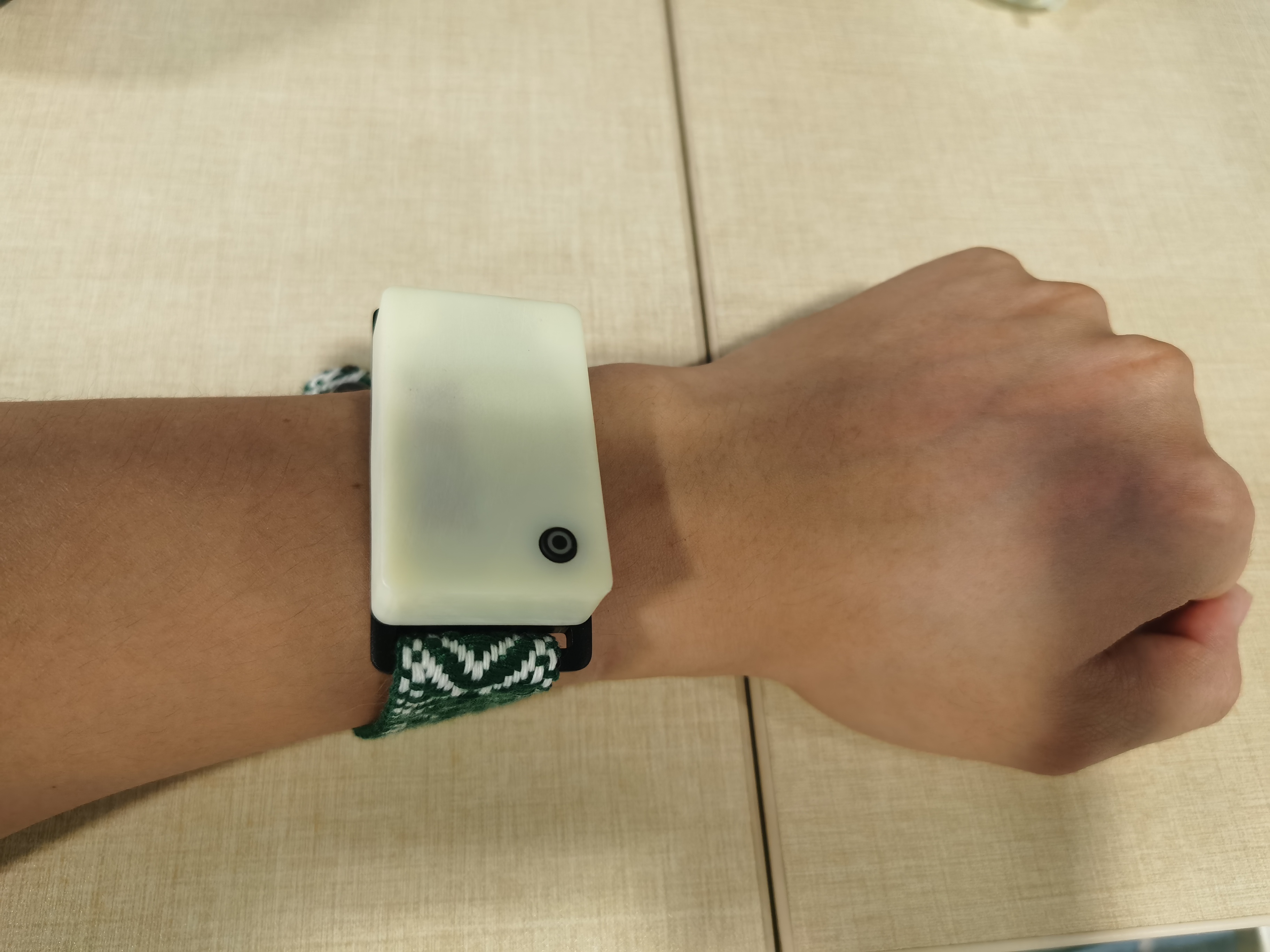}
		\caption{Wearable device}
		\label{fig:device}
         
	\end{minipage}
 
\end{figure}

\subsection{Data Collection}
As there are no publicly available multi-sensor wearable detection data and corresponding ECG signal data, we collect a dataset ourselves to implement and evaluate the system. We wear the wearable device on the subject's wrist and use the ECG device for simultaneous acquisition of ground truth ECG signal data. In order to adapt the model to a variety of office scenarios, we have each participant perform three actions, sitting, typing, and arm swinging, to simulate different situations that would occur in daily office scenarios where the arm enters the field, which would allow our model to better achieve health monitoring in office scenarios. We recruited 15 subjects for the experiment. All subjects report no heart disease and are in good health. Each subject collects data for 10 minutes in each scenario. After collecting the data, we uniformly resample the data to 50Hz because the data sampling frequency is different for different devices.

\section{Evaluation} \label{sec:secevaluation}
In this section, we first evaluate the overall performance of \name and compare it with existing wearable ECG monitoring methods \cite{cao2022guard} and linear ICA methods \cite{hyvarinen2000independent}. We also compare it with the existing wearable COTS device with ECG monitoring capabilities. Then, we will conduct sufficient experiments to verify the performance of our proposed Deep-ICA module and multimodal fusion module. Finally, we analyze the effect of different experimental settings on the results.

\subsection{Evaluation Metrics}

We use the Pearson correlation coefficient (CC), a statistic that measures the strength and direction of the linear relationship between two continuous variables \cite{cohen2009pearson}. It can take values between -1 and 1. The cosine distance (COS) is a measure of the similarity between two vectors, commonly used in areas such as text mining and information retrieval. Mathematically, the COS measures the angle between two vectors and can take values from 0 to 1, where 1 means that the two vectors are completely similar and 0 means that the two vectors are completely dissimilar.  Root Mean Square Error (RMSE) is a commonly used metric for evaluating regression models that measure the degree of discrepancy between the observed values and the predicted values of the regression model. The smaller the RMSE, the better.

Since the waveforms of the COTS device do not quite match the waveform characteristics of our ground truth in the experiments compared to the device, it is not reasonable to use the above evaluation metrics for evaluation. Therefore, we adopt the method \cite{foster2020inertial} of similar evaluation metrics for evaluation. Specifically, we use the precise detection and alignment of R-peaks to the metric. For the peak alignment algorithms, we set the parameters as a distance threshold of 0.25 times the sampling rate and a tolerance of one error window. To evaluate the models used, we classify peak alignments as true positives (TP), misaligned true R-peaks as false negatives (FN), and misaligned predicted R-peaks as false positives (FP). We calculate precision, recall, and F1 from these classifications.

\subsection{Overall Performance}

\begin{figure}[t]
\centering  
\subfigure[Ground truth ECG]{
\label{fig:ground}
\includegraphics[width=0.47\linewidth,trim={ 0 0 1.3cm 0},clip]{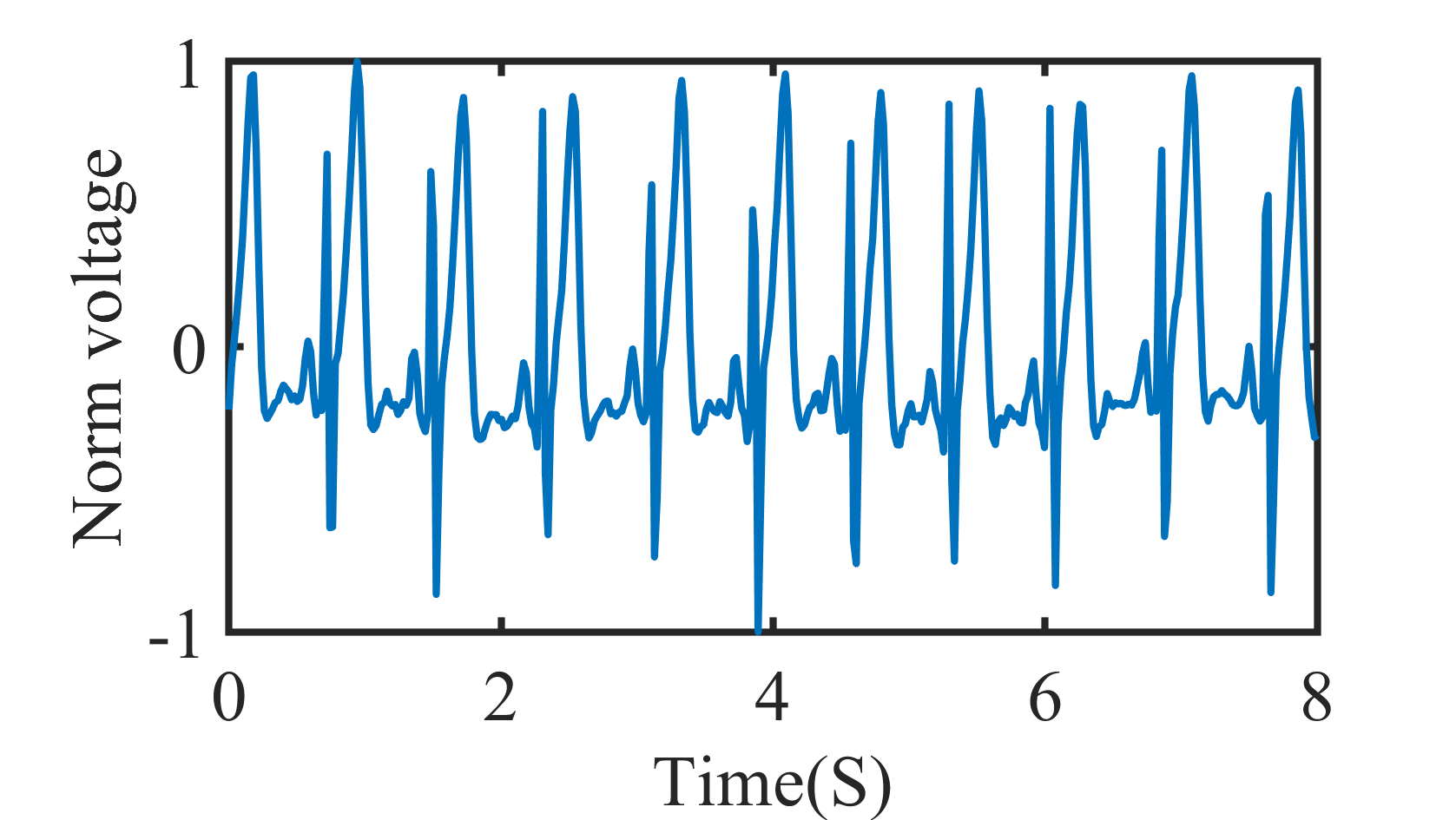}}
\subfigure[Generated ECG]{
\label{fig:genecg}
\includegraphics[width=0.47\linewidth,trim={ 0 0 1.3cm 0},clip]{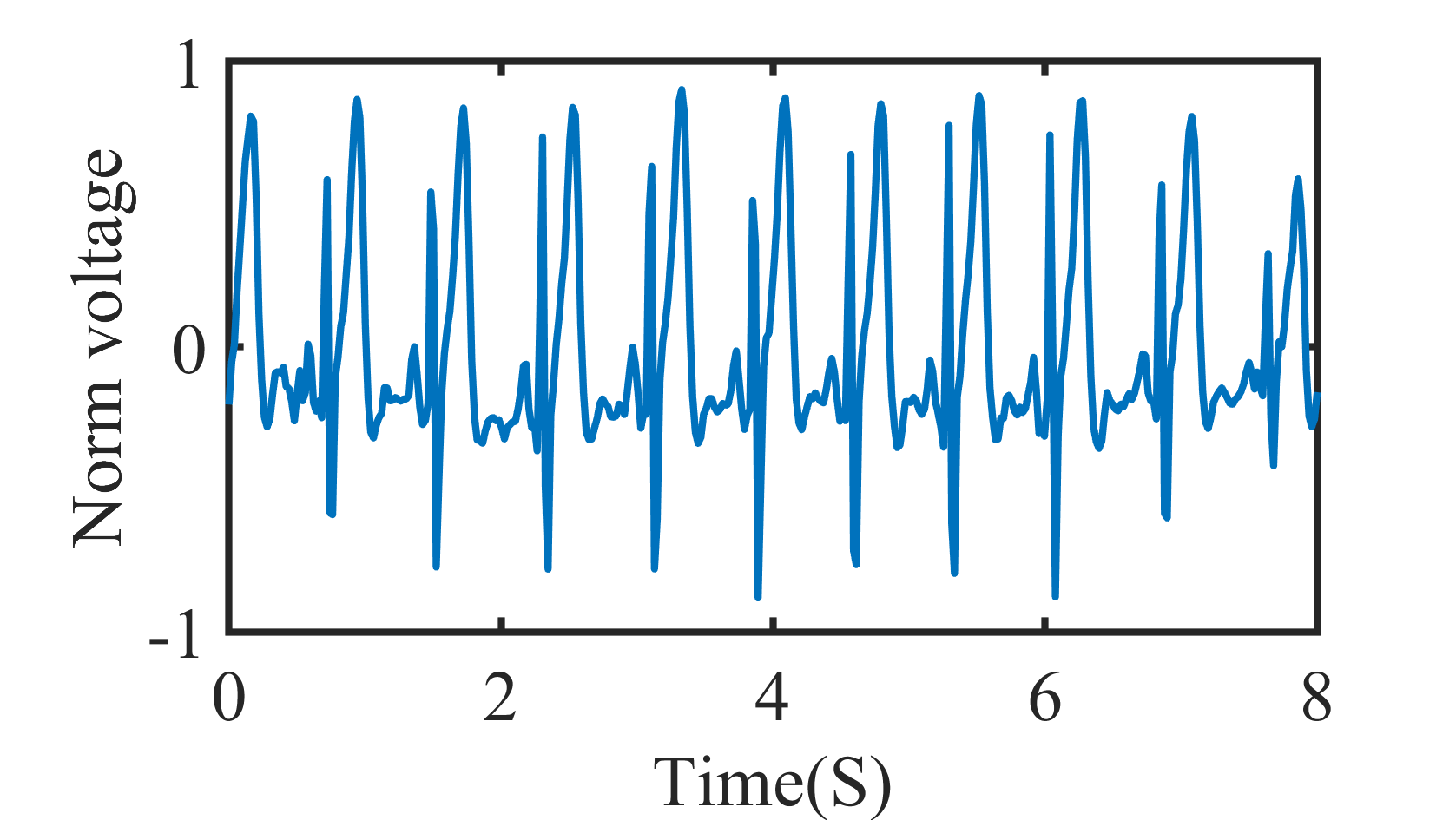}}
\caption{Comparison of ground truth with the generated ECG signal}
\label{fig:comp genecg}
\end{figure}

First, we compare the performance with VibCardiogram \cite{cao2022guard}, which is an IMU-based wearable ECG monitoring method. In addition, we have included a single-mode ablation experiment of the \name method using only IMU, which allows us to show the respective differences more intuitively. We use IMU to indicate that only a single modality of IMU is used for the experiments. We then also perform a comparison experiment with linear ICA. From Table \ref{tab:Vib_compa}, we can observe that the results of IMU ablation and VibCardiogram are not very good when using a single modality alone. This suggests that it is difficult to achieve cardiac monitoring in office scenarios when relying solely on single-modality data. However, we can see significant improvements in various evaluation metrics when utilizing multi-modal data. This proves the validity of our multimodal approach. In addition, we can also observe that denoising the original data using only linear ICA has a negative effect on performance, which indicates that the ordinary linear ICA method cannot solve the noise problem well.  Fig. \ref{fig:comp genecg} shows the comparison between the model-generated ECG and ground-truth, and we can see that the generated ECG waveforms are able to reconstruct the features in the original ECG relatively well.

\begin{table}
    \centering
    \caption{Comparing \name to IMU Ablation and Other Network.}
    \label{tab:Vib_compa}
    \begin{tabular}{lccc}
        \toprule
        \textbf{Method}& \textbf{CC}& \textbf{COS}& \textbf{RMSE}\\ 
        \midrule
        VibCardiogram &0.5218&0.5541&0.1840\\
        FastICA &0.4513&0.4929&0.1820\\
        IMU &0.5269&0.5698&0.1599 \\
        \textbf{\name} & \textbf{0.8235}& \textbf{0.8446}& \textbf{0.0940}\\
    \bottomrule
\end{tabular}
\end{table}

\subsection{Scenario Evaluation}
\begin{figure}[t]
\centering  
\subfigure[CC]{
\label{fig:sence_CC}
\includegraphics[width=0.31\linewidth,trim={ 0 0 1.3cm 0},clip]{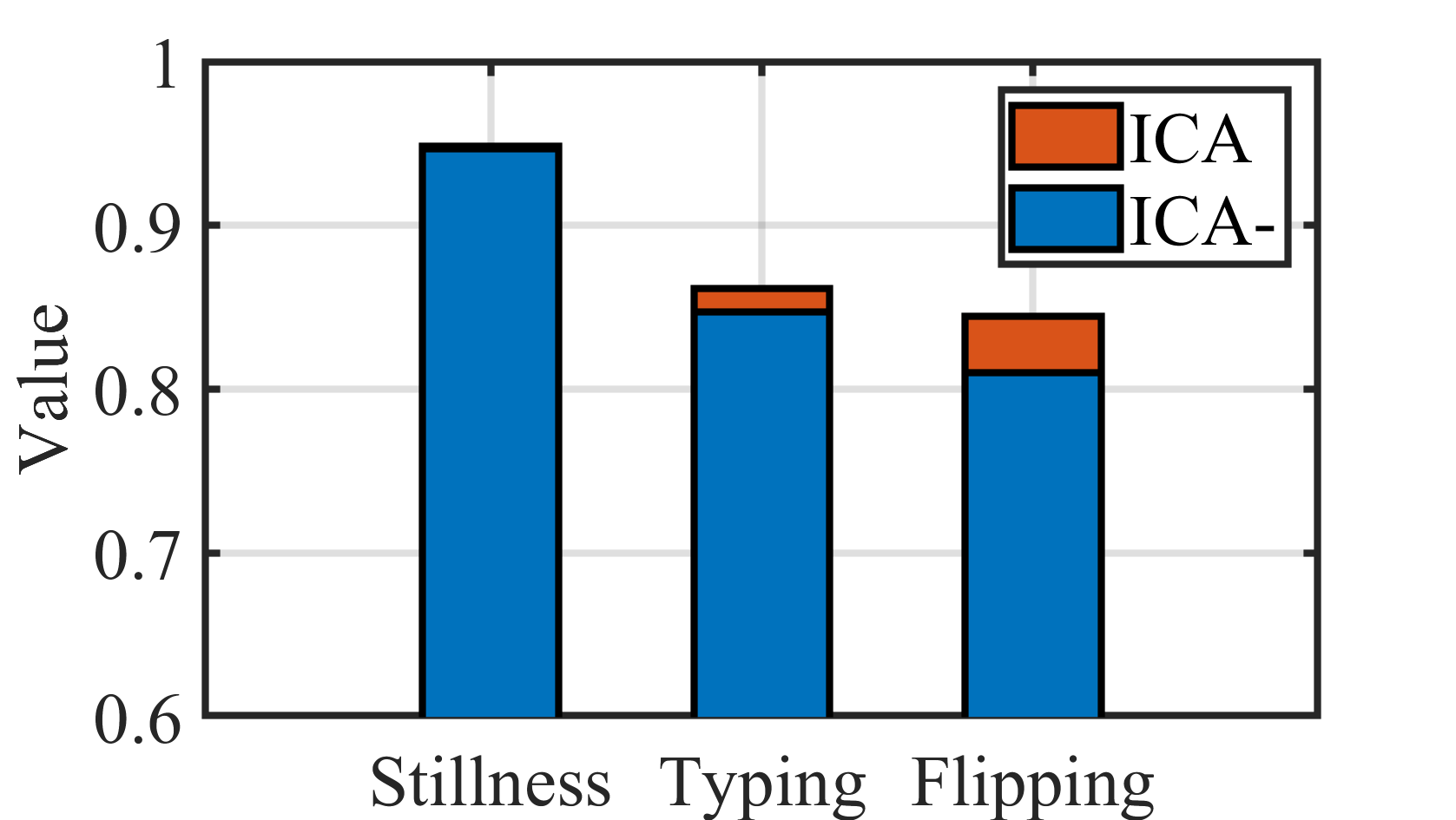}}
\subfigure[COS]{
\label{fig:sence_COS}
\includegraphics[width=0.31\linewidth,trim={ 0 0 1.3cm 0},clip]{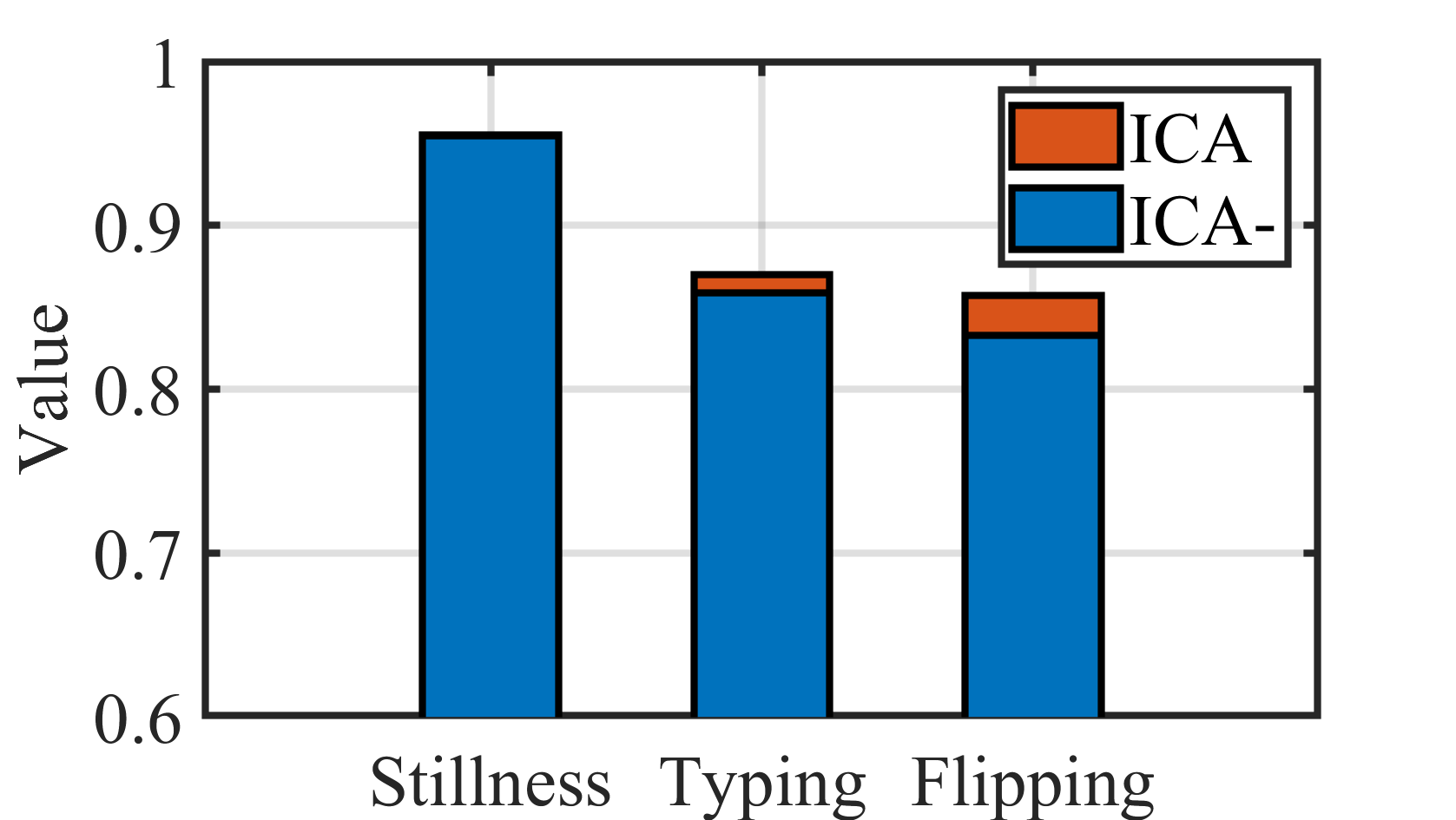}}
\subfigure[RMSE]{
\label{fig:sence_RMSE}
\includegraphics[width=0.31\linewidth,trim={ 0 0 1.3cm 0},clip]{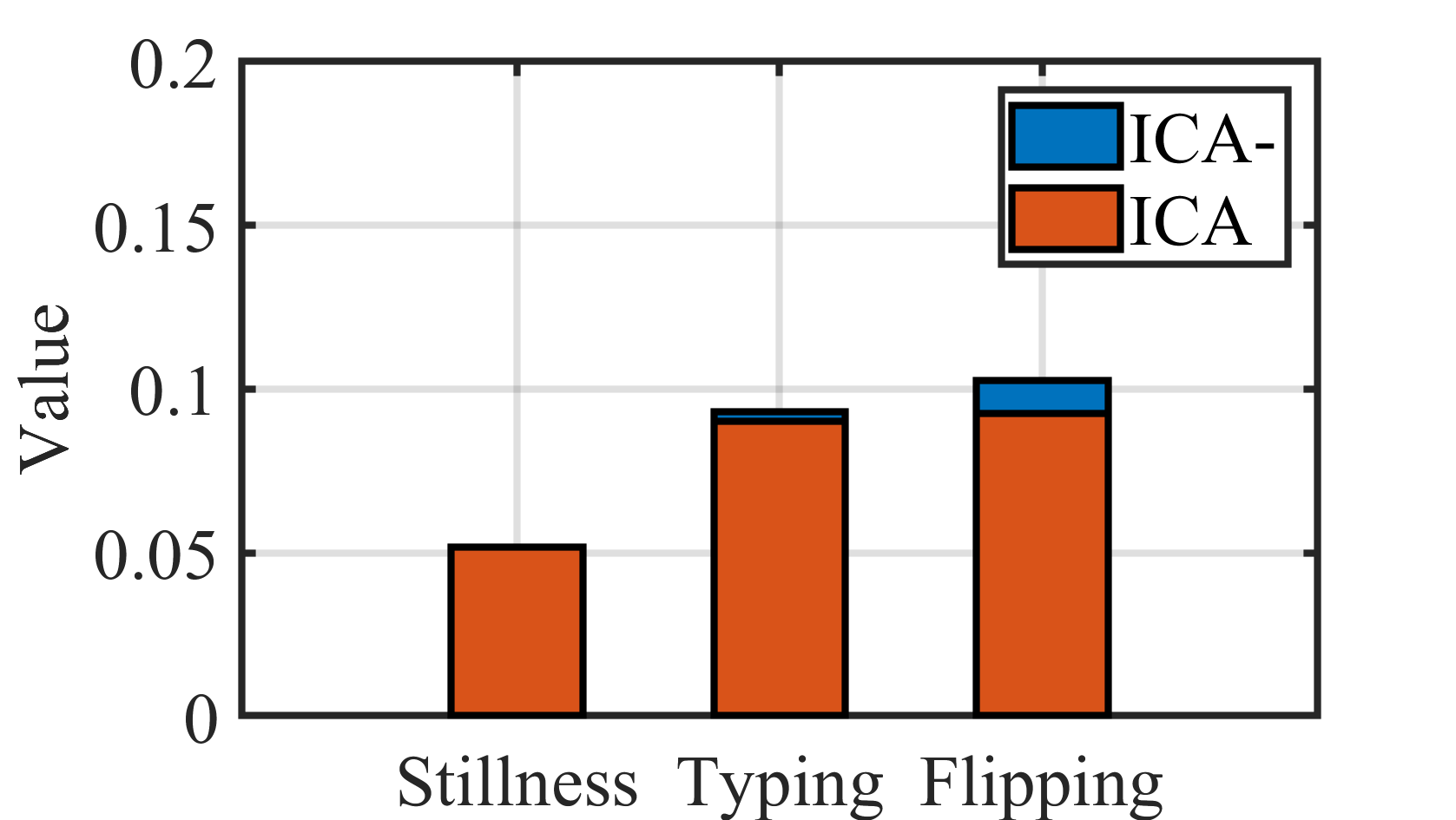}}
\caption{Comparison of ICA Modules in Different Scenarios. \textbf{ICA-}: Ablation experiments without ICA module, if no ICA indicates that ICA module was used. Higher values of CC and COS are better, and lower values of RMSE are better.}
\label{fig:sence_comp}
\end{figure}

We also perform experiments in the proposed three scenarios of stillness, typing, and flipping. From Fig. \ref{fig:sence_comp}, we can see that the stillness scenario has the best results even more than the mixed scenario, because in this case there is no interference from the arm movement, and the model can better extract the characteristics of the ECG signal waveform from the sensor data. Then we can see that the typing scene has better results compared to the flipping scene, we have analyzed this phenomenon and found that the movements in the typing scene are more regular and the amplitude of the movements is smaller compared to the movements in the flipping scene. So the data in the typing scene will have a better effect on the model.

We also perform experimental comparisons of a single modality in different scenarios. The results of the realization are shown in Fig. \ref{fig:sence_comp}. As we analyzed earlier, IMU is more sensitive to small vibrations, so IMU will be better than radar in still scenes. In a motion scene, the IMU sensor collects data from three directional axes, so the movement of the arm has a greater effect on the IMU sensor. Due to the difference in the mechanism, the relative distance between the radar sensor and the blood vessel is unchanged, so the movement of the arm has less effect on the radar, and we can also see from the experiments that the effect of the radar is better than that of the IMU in all motion scenarios. We also observe that performance improves to varying degrees in all three scenarios after using data from multiple modalities, demonstrating the effectiveness of our proposed multimodal fusion module.

\subsection{COTS Comparison}
To demonstrate the effectiveness of our approach, we compare the performance with that of existing commercial devices. The off-the-shelf watch we used is the Huawei Watch 4. This watch is NMPA medically certified and can give the user a single lead ECG with a sampling rate of 500Hz. The measurement principle is that the fingers of the other hand must touch the watch, so that the fingers, the watch, and the heart form a complete closed loop. When the heart beats, the heart muscle cells become excited and generate very small electrical signals on the surface of the skin. The watch amplifies and records these electrical signals. Since the watch requires two hands when in use, the movements in our typing scenario cannot collect data in this case, so we use the two scenarios of stillness and flipping for our experiments. since the waveform of the watch has a large difference from the waveform of our ground truth, we do not use this writing evaluation index mentioned above, and we follow the R-peak accuracy as the evaluation index, specifically, we set an error interval of 0.25 times the sampling rate, if the position of a single R-peak in the ECG waveform of the watch is the same as the position of the R-peak in the ground-truth, we consider it as a positive example, and if it is not within this error interval, we consider it as a negative example. We use COTS to represent the watch. From Table \ref{tab:COTS_compa}, we can observe that in the stillness scenario, both our method and COTS have a good result, and the precision can reach more than 90\%. However, in the flipping scenario, we can see that the effect of COTS is greatly weakened, but our method can still maintain an accuracy of 86\%.

\begin{table}[t]
    \centering
    \caption{Comparison of the results of our method with COTS equipment.}
    \label{tab:COTS_compa}
    \begin{tabular}{lccc}
        \toprule
        \textbf{Method}& \textbf{Precision}& \textbf{Recall}& \textbf{F1}\\ 
        \midrule
        COTS Flipping &0.6962& 0.7143& 0.7051  \\ 
        \name Flipping &0.8645&0.8603&0.8624 \\ 
        COTS Stillness &0.9429& 0.9296& 0.9362 \\ 
        \name Stillness &0.9414&0.9424&0.9419 \\  
    \bottomrule
\end{tabular}
\end{table}

\subsection{Ablation Study}

\begin{figure}[t]
\centering  
\subfigure[CC]{
\label{fig:ICA_CC}
\includegraphics[width=0.31\linewidth,trim={ 0 0 1.3cm 0},clip]{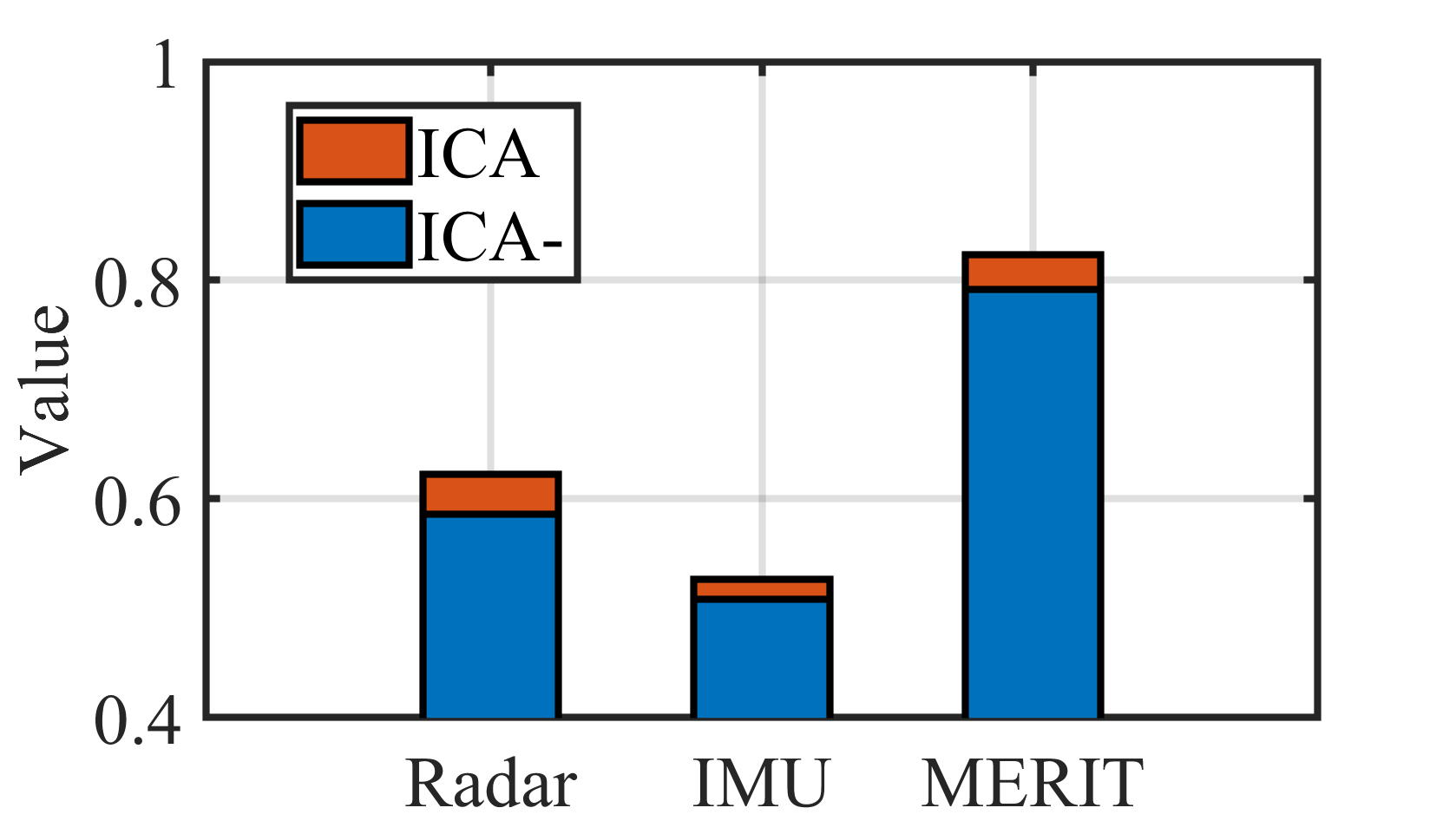}}
\subfigure[COS]{
\label{fig:ICA_COS}
\includegraphics[width=0.31\linewidth,trim={  0 0 1.3cm 0},clip]{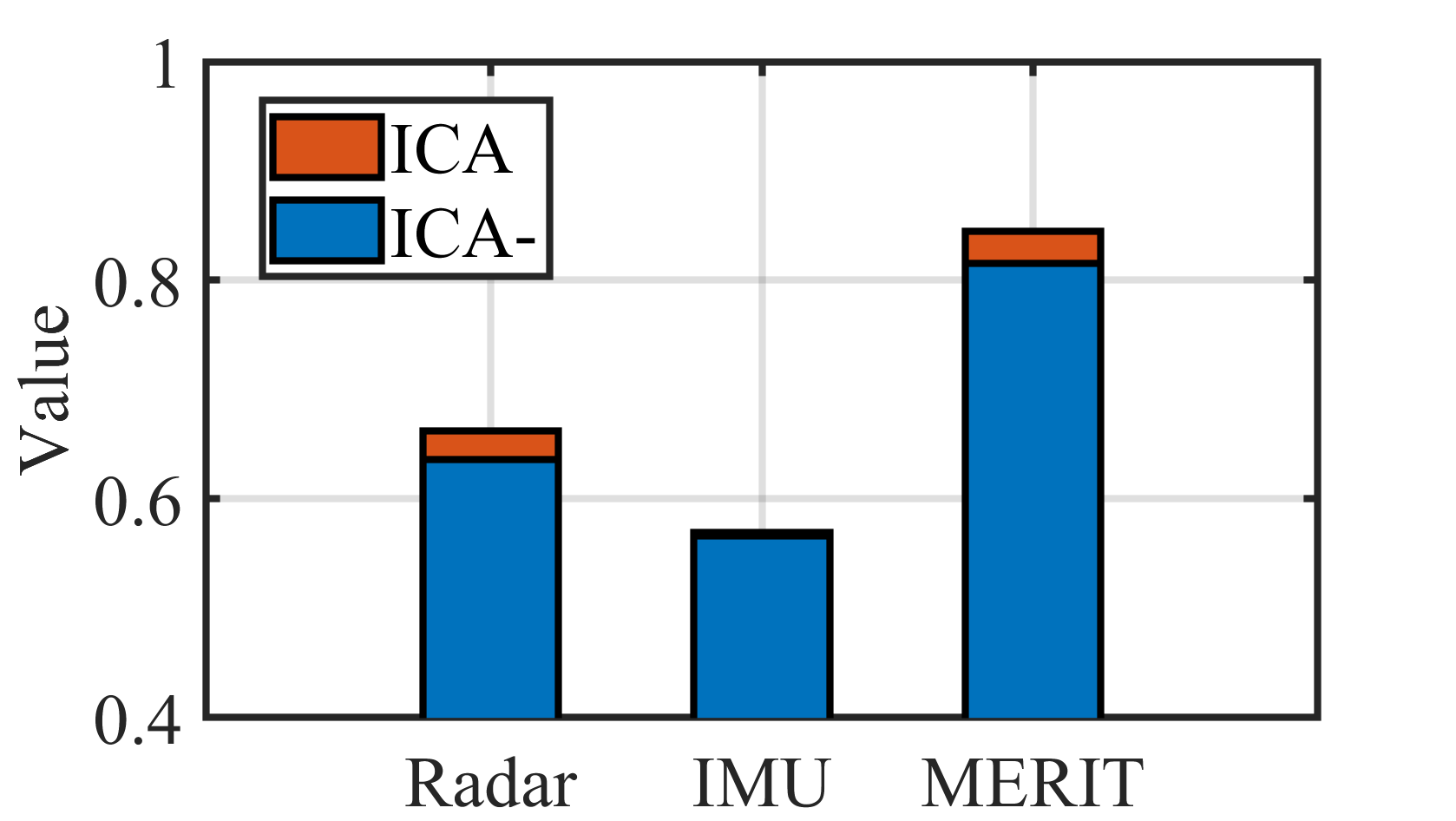}}
\subfigure[RMSE]{
\label{fig:ICA_RMSE}
\includegraphics[width=0.31\linewidth,trim={ 0 0 1.3cm 0},clip]{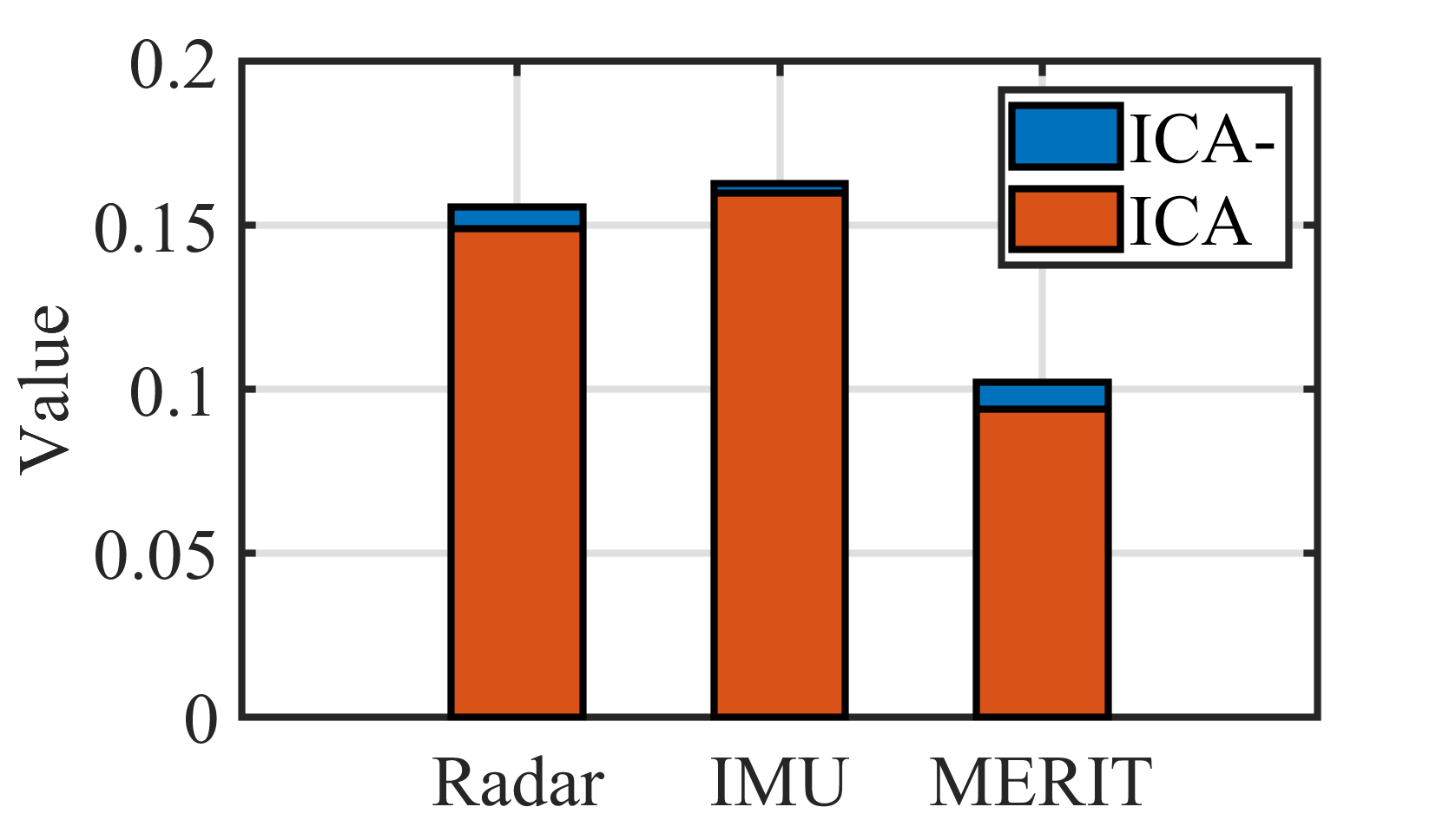}}
\caption{Ablation experiments with different modules and ICA modules. \textbf{ICA-}: Ablation experiments without ICA module, if no ICA indicates that ICA module was used. Higher values of CC and COS are better, and lower values of RMSE are better.}
\label{fig:ICA_Ablation}
\end{figure}

To prove the effectiveness of our proposed two modules, we do a complete modal ablation experiment on the multimodal side. We also add the ICA module ablation experiment while performing the different modes of ablation. In this way, the effects of different modes and ICA modules on the model can be fully demonstrated. The Radar and IMU denote the single-mode ablation experiments, and ICA- denotes the ablation experiments without the ICA module. From Fig. \ref{fig:ICA_Ablation} we can observe that in the case of using only, Radar and IMU single modes, the results have decreased to different degrees compared to the multimodal condition. This further validates the radar and IMU scenarios we analyzed earlier. Both radar and IMU have their own scenarios that they are good at dealing with, and the results when we use only one sensor modality to face the data from all the scenarios are not good, but the results are improved after using both sensor modalities. Also, we can see that after using the ICA method it is in the case of both single modes that the final results are improved. This also shows that our proposed method is better able to alleviate the effects of motion.

\subsection{Impact Factors}
\subsubsection{Experimental Parameters}

\begin{figure}[t]
\centering  
\subfigure[BatchSize.]{
\label{fig:Batch Size}
\includegraphics[width=0.47\linewidth,trim={ 0 0 1.3cm 0},clip]{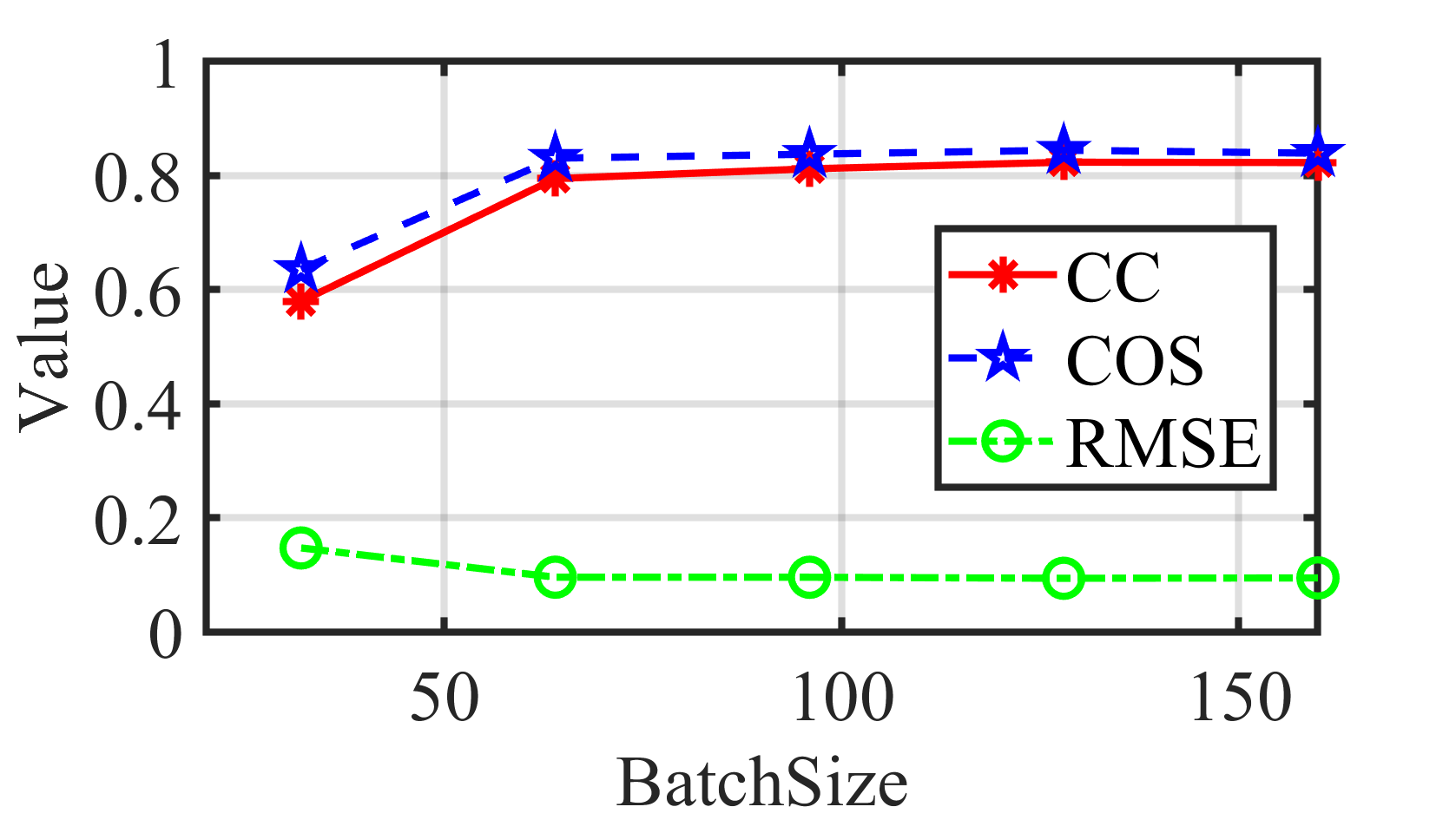}}
\subfigure[$\alpha$ parameter.]{
\label{fig:alpha}
\includegraphics[width=0.47\linewidth,trim={ 0 0 1.3cm 0},clip]{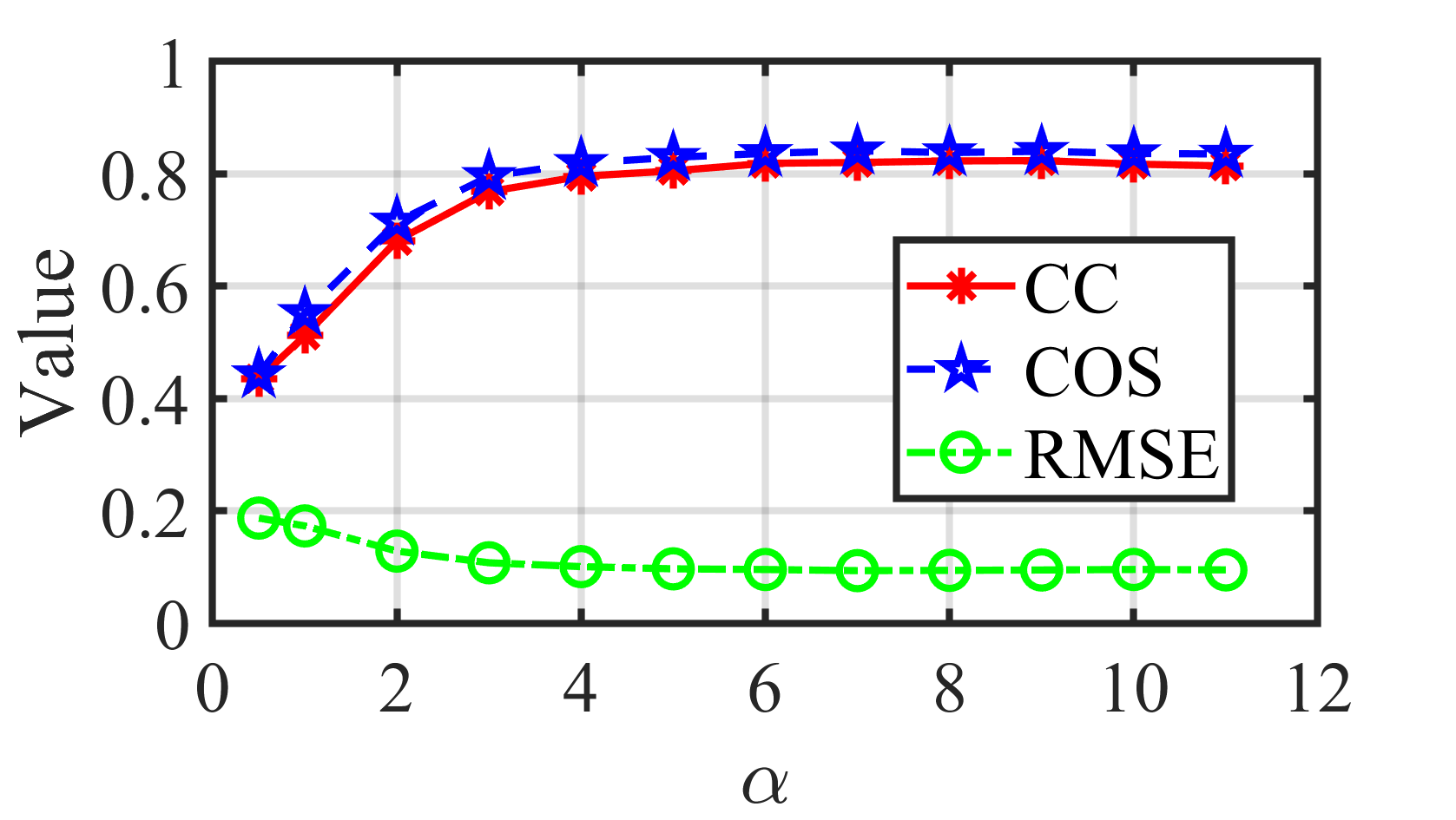}}
\caption{Impact of BatchSize and $\alpha$ parameter.}
\label{fig:Batch alpha}
\end{figure}

We perform an experimental analysis of the BatchSize and the $\alpha$ parameter in Equation \ref{eq:ganloss}. As shown in Fig. \ref{fig:Batch Size}. We find that too small a BatchSize has a large impact on performance, and we analyze that smaller BatchSize gradient estimates may have a large variance, leading to a more unstable training process. The performance tends to stabilize when the BatchSize is gradually increased. We end up setting the BatchSize to 128. Fig. \ref{fig:alpha} shows the effect of different values of $\alpha$ on performance. We find that a value of $\alpha$ of 9 has the best performance.

\subsubsection{Deep-ICA module Input}

We also explore the effect of the ICA module on the antenna dimension and the range-bin dimension. Ant indicates that the ICA module is performed in the antenna dimension, and Idx indicates that the ICA module is performed in the range-bin dimension. Table \ref{tab:ant_idx_compa} shows that the effect in the antenna dimension will be better than the effect in the range-bin dimension. We have analyzed this situation because the range of human wrist pulse blood vessel beating is relatively small and only one of the target bins in range-bin can detect the blood vessel changes well. If we use the target bin as well as the next two bins, we find that the next two bins do not contain much information about the blood vessel movement, and these two bins are not very helpful for the ICA module. But the antennas are different, we select the target bin from different antennas so that the bins selected from different antennas can contain the information on the blood vessel movement, so the result will be better.

\begin{table}[t!]
    \centering
    \caption{Effect of using different antennas and different bins on ICA modules.}
    \label{tab:ant_idx_compa}
    \begin{tabular}{lccc}
        \toprule
        \textbf{Method}& \textbf{CC}& \textbf{COS}& \textbf{RMSE}\\ 
        \midrule
        Idx &0.7308&0.7615&0.1144 \\ 
        Ant &0.8235 &0.8446 &0.0940  \\ 
    \bottomrule
\end{tabular}
\end{table}

\section{Conclusion}
In this paper, we propose a wearable method for vital signs detection \name that enables human health monitoring by capturing wrist vascular vibrations and reconstructing them to ECG signals. We have designed two modules for this purpose, the Deep-ICA module, which uses non-linear ICA to reduce the effect of arm noise on performance, and the multimodal fusion module, which combines the characteristics of both IMU and radar sensors to better reconstruct the ECG signal. We have conducted sufficient experiments to prove the effectiveness of our proposed method. For future directions, it is worth exploring how to deploy larger-scale models such as large language models on resource-constrained wearables~\cite{lin2023pushing,fang2024automated,lin2024splitlora,qiu2024ifvit}, and how to collaboratively train models while protecting user privacy~\cite{zhang2024satfed,zheng2023autofed,lin2024efficient,zhang2024fedac,lin2024adaptsfl,lin2024split,lyu2023optimal,lin2023fedsn}.

\bibliographystyle{IEEEtran}
\bibliography{Newbib}

\end{sloppypar}
\end{document}